\documentclass[aip,amsmath,amssymb,reprint]{revtex4-1}

\usepackage{graphicx}
\usepackage{amsmath}
\usepackage{dcolumn}
\usepackage{color}
\usepackage{epstopdf}
\usepackage{subfigure}

\newcommand{\B}{{\bf B}}

\newcommand{\Bt}{{\tilde B}}

\newcommand{\vt}{{\tilde v}}

\newcommand{\ki}{{\bf k}}

\newcommand{\e}{{\bf e}}

\newcommand{\x}{{\boldsymbol \xi}}

\newcommand{\wb}{\bar \omega}

\usepackage{titlesec}
\titlespacing*{\section} {0pt}{1.5ex}{1.5ex}
\begin{document}


\title{A generalized effective potential for differentially rotating plasmas} 



\author{Fatima Ebrahimi}
\email[]{ebrahimi@princeton.edu}
\affiliation{Princeton Plasma Physics Laboratory,and the Department of Astrophysical Sciences, Princeton University, New Jersey 08540, USA}

\author{Alexander Haywood}
\email[]{ahaywood@princeton.edu}
\affiliation{Department of Mechanical and Aerospace Engineering \\
Princeton University, New Jersey 08544, USA}


\date{\today}

\begin{abstract}
Global stability of differentially rotating plasma is investigated using a generalized effective potential. We first, for a current-free system, obtain a general form of an effective potential in terms of the free energies of global curvature and gradients of rotation for non-axisymmetric disturbances. We then examine the stability of differentially rotating disks for several rotation profiles and present the associated effective potential for the onset of these instabilities in the MHD regime. In particular, results for global axisymmetric magnetorotational instability (MRI) as well as local and global non-axisymmetric modes are presented. The latter constitute two distinct non-axisymmetric modes, a high frequency local MRI and a global low-frequency non-axisymmetric mode (the magneto-curvature mode, introduced in Ebrahimi\&Pharr, ApJ 2022), confined either between two Alfv\'enic resonances or an Alfv\'enic resonance and a boundary.  

\end{abstract}


\maketitle 

\section{Introduction}
Global axisymmetric magnetorotational instability (MRI) ~\cite{velikhov1959, chandrasekhar60, balbus1991powerful} is believed to be a potential driver of the turbulence in astrophysical disks that lead to the occurrence of the accretion process.   This requires a weak magnetic field in a differentially rotating disk. Global non-axisymmetric disturbances could, however, be another important driver of turbulence, as they could source dynamo fields.~\cite{rincon2007, Ebrahimi_2009,ebrahimi_blackman_2016,bhat_ebrahimi_blackman_2016} Although non-axisymmetric instabilities are less studied theoretically, they have been investigated in laboratory experiments~\cite{sisan2004experimental,spence2012free,caspary2018effects,choi2019nonaxisymmetric,mishra2022}. The observation of a novel exponentially growing non-axisymmetric mode in the MRI experiment at PPPL~\cite{wang2022identification}  has, in particular, generated a new interest for better understanding of non-axisymmetric instabilities in the differentially rotating systems.  

Hydrodynamically stable flows with $\Omega(r) \sim r^{-q}$, $q<2$ (according to Rayleigh criteria) are MHD unstable in the presence of uniform axial magnetic field, as has also been shown in the laboratory~\cite{wang2022observation}. Purely growing axisymmetric MRI modes can be understood via local treatments~\cite{balbus1991powerful}. However, the physics of non-axisymmetric instabilities is more complex and require global treatment. The oscillatory behavior (not purely growing) of these modes, as well as the possible existence of resonances in the domain, could contribute to the rich physics of non-axisymmetric instabilities in rotating systems. Here, we extend the physics understanding of these modes using a global linear analysis by exploiting the characteristics of a generalized effective potential. 


The interactions between waves of
positive and negative action is known to lead to an instability.~\cite{cairns1979} In the context of astrophysical and hydrodynamical systems, sound, surface and Rossby waves can give rise to global non-axisymmetric perturbations (see for example ~\cite{papaloizou1984,goldreich1986,goodman1987stability,  
 lovelace1999rossby,glatzel1987,ono2016}). In the MHD model with the presence of a magnetic field B,
an Alfv\'enic resonance, where the magnitude of the Doppler-shifted wave frequency is equal to the Alfv\'en frequency, could additionally cause the onset of non-axisymmetric modes. For non-axisymmetric MRI modes, localized structures were found to be confined between the Alfv\'nic resonant points in Cartesian geometry~\cite{matsumoto1995magnetic}, in cylindrical shear flows~\cite{Ogilvie_1996} and  in the compressible limit~\cite{goedbloed2022} at large axial ($k$) and azimuthal ($m$) mode numbers.
Additionally, for the same B and wave numbers ($m$, $k$), two distinct modes, at two different frequencies were found~\cite{ebrahimi2022}, 1) a local MRI at high frequency (mostly confined between the two Alfv\'enic points, in the inner mode close to the inner boundary) and 2) a low-frequency mode either confined between the Alfv\'enic resonances (the outer mode close to the outer boundary) or confined between an Alfv\'enic resonance and a boundary. The latter low-frequency mode, the so called magneto-curvature~\cite{ebrahimi2022} (or curvature MRI) mode, is a global non-axisymmetric mode due to global curvature and differential rotation and persists at stronger magnetic field.

Here, we find that given the free energy (i.e. first and second derivatives of global rotation) either a potential well is formed and extended globally between two Alfv\'enic points (Alfv\'enic propagating region) to confine a non-axisymmetric mode or a negative potential changes sign between one resonance and a boundary. 
Below, we first present the basic equations and the derivation of the effective potential. Numerical solutions will follow.

\section{Basic equations and the derivation of effective potential}
We revisit the linear global stability analysis obtained in Ref.~\cite{ebrahimi2022}  In the ideal limit and with cylindrical coordinates,  the velocity and magnetic perturbations can be expressed in terms of the displacement vector $\x(r, \phi, z, t) = [\xi_r(r),\xi_{\phi}(r),\xi_z(r)]\exp{i(m\phi+kz - \omega t)}$  \cite{Chandrasekhar_2006}, as $ {\bf \vt} = -i\wb\x - r\frac{\partial \Omega}{\partial r} \xi_r \e_\phi$ and ${\bf \Bt} = i\left(\ki\cdot\B\right)\x+\frac{2 B_\phi}{r}\xi_r \e_\phi$, respectively. Using the incompressibility,
   $ \nabla\cdot\x = \frac{1}{r}\frac{\partial}{\partial r}\left( r\xi_r\right) + \frac{im}{r}\xi_\phi + i k \xi_z = 0$, we obtain the components of the linearized momentum equation,
   \begin{multline}
 \left(-\wb^2 + \omega_A^2 + \omega_s^2 + \omega_c^2 \delta_c\right) \xi_r +\left[i\omega_A\omega_c \delta_c+ 2i\wb \Omega(r)\right] \xi_{\phi} \\= - \frac{\partial \widetilde P}{\partial r}\label{eq:mri1}
 \end{multline}
 \begin{multline}
\left(-\wb^2 + \omega_A^2\right) \xi_{\phi}  -i \left[2\wb \Omega(r) + \omega_A \omega_c \delta_c\right] \xi_{r} \\= -\frac{im \widetilde P}{r}
\label{eq:mri2}
\end{multline}
\begin{equation}
\left(-\wb^2 + \omega_A^2\right) \xi_{z} = -i k_z \widetilde P,
\label{eq:mri3}
\end{equation}

where $\omega_A \equiv \frac{\ki\cdot\B_0}{\sqrt{\mu_0 \rho}} = \frac{1}{\sqrt{\mu_0\rho}}\left( k_z B_z + \frac{m}{r}B_\phi \right)$
$\omega_c \equiv \frac{2B_\phi}{r\sqrt{\mu_0\rho}}$,  $\omega_s^2 \equiv \frac{\partial \Omega^2}{\partial \ln r}$,
$\wb = \omega - m \Omega(r)$, $\omega = \omega_r +i\gamma$,  and $\widetilde P \equiv \frac{1}{\rho}\left(\widetilde{p} + \frac{1}{\mu_0}{\bf \Bt}\cdot \B_{0}\right)$. 
By combining (\ref{eq:mri1}-\ref{eq:mri3}), an ordinary differential equation was obtained~\cite{ebrahimi2022}, 
\begin{multline}
\label{eq:ode}
    \left[ \frac{\Gamma(x)}{Q(x)}xu'\right]' + u \left[  m \left(\frac{\Omega\wb + \frac{1}{2}\omega_A\omega_c}{Q(x)}\right)'\right. \\ \left. - \frac{(\wb^2 - \omega_A^2 - \omega_s^2 - \omega_c^2)}{4x} + \frac{\left(\Omega\wb + \frac{1}{2}\omega_A\omega_c\right)^2}{Q(x)}
     \frac{k^2}{\Gamma(x)}\right] = 0
\end{multline}

Here we have defined $u \equiv r\xi_r $ and $x \equiv r^2$, and used 
\begin{multline}
\Gamma (x) = \wb^2-\omega_A^2, 
\quad Q(x) = k^2x +m^2\\
\end{multline}
With the change of variable $\Psi =A u$, where $A = \sqrt{\frac{\Gamma(x)}{Q(x)} x}$, we rewrite Eq.~\ref{eq:ode} as
\begin{equation}
\Psi^{\prime\prime} (x) - U(x,\omega,m) \Psi(x)=0 , 
\label{eq:schrodinger}
\end{equation}
and, 
\begin{equation}
U(x,\omega,m) = \frac{A^{\prime\prime}}{A} -\frac{C}{A^2}
\label{eq:gpotential}
\end{equation}
where, 
\begin{multline}
\frac{A^{\prime\prime}}{A}=\frac{\left[x \Gamma^{\prime\prime}(x) + 2 \Gamma^{\prime}(x)\right]}{2x\Gamma(x)} -\left[\frac{\left(x\Gamma^{\prime}(x)+\Gamma(x)\right)}{2 x\Gamma(x)}\right]^2\\
-k^2 \frac{\left[x\Gamma^{\prime}(x)+\Gamma(x)\right]}{2x Q(x) \Gamma(x)}+\frac{3k^4}{4Q^2(x)}\\
\frac{C}{A^2}= \frac{Q(x)}{x\Gamma(x)}\left[  m \left(\frac{\Omega\wb + \frac{1}{2}\omega_A\omega_c}{Q(x)}\right)'\right. \\ \left. - \frac{(\wb^2 - \omega_A^2 - \omega_s^2 - \omega_c^2)}{4x} + \frac{\left(\Omega\wb + \frac{1}{2}\omega_A\omega_c\right)^2}{Q(x)}
     \frac{k^2}{\Gamma(x)}\right]
     \label{eq:terms}
\end{multline}

Equation~\ref{eq:schrodinger} is equivalent to a Schrodinger-like  equation, where an effective potential $ U(x,\omega,m)$ is expressed in terms of differential rotation and magnetic fields (in a current-free system). 
Equation~\ref{eq:gpotential} is a generalized form of the effective potential for non-axisymmetric disturbances ($m\neq0$), which contains free energies from (hydrodynamics or MHD) flows  with their full radial  extent, as well as curvature of fields and space. Here, we examine the characteristics of this potential for various flow profiles and the associated instabilities.   

For axisymmetric modes (m=0) (with a purely growing mode, i.e. real $\omega^2=-\gamma^2$)~\cite{chandrasekhar60} in purely axial field ($\textbf{B}=B_z \hat{z}$, $\wb=\omega$, $A^{\prime\prime}=0$), the effective potential reduces to


\begin{equation}
U(x,\omega,0) = \frac{k^2}{4x}\left[1 -\frac{4 x \Omega (x) \Omega(x)^{\prime}}{\omega^2-\omega_A^2} -\frac{4 \Omega(x)^2 \omega ^2}{(\omega^2-\omega_A^2)^2}\right]
\label{eq:m=0_potentia}
\end{equation}

Here, the zero energy solutions for the frequency dependent potential, $U(x, \omega)$, provides key information about the global stability. To guarantee an unstable global mode (which is non-monotonic to meet the boundaries), the effective potential should be negative somewhere in the domain ($U(x, \omega)<0$)~\cite{pino2008global}. From this condition ($U(x,0) <0$, for marginal stability $\omega^2 =0$), we can immediately arrive to the condition for the axisymmetric MRI instability  $\omega_A^2 < - \omega_s^2$ [ 
$=\frac{\partial \Omega(r)^2}{\partial \ln r}= 4x\Omega(x)\Omega(x)^{\prime}]$.

\begin{figure}
\vspace{-8mm}
    \centering

    \includegraphics[width=0.9\linewidth]{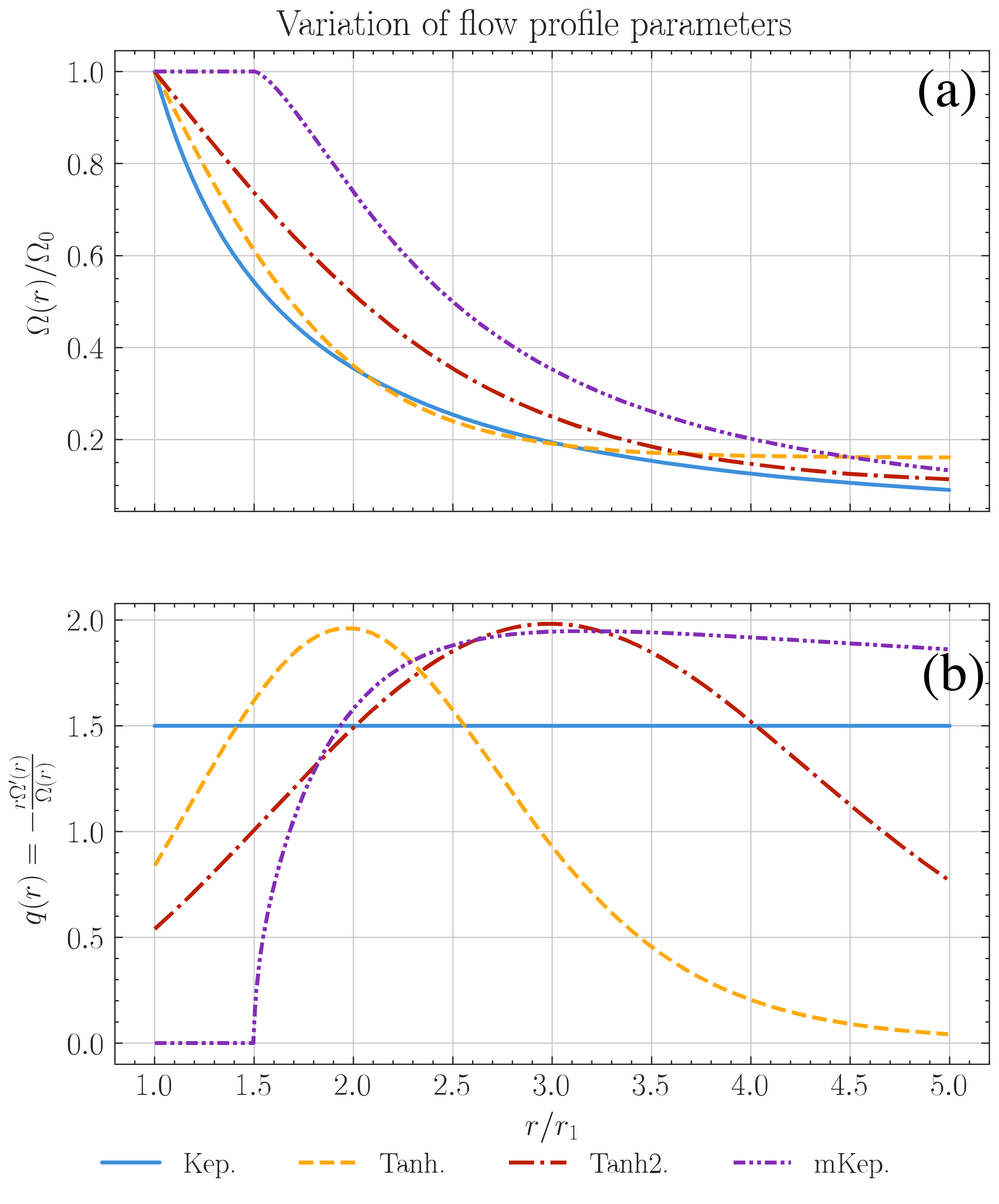} 
  \caption{Variation of rotation and $q(r) = \frac{r\Omega'(r)}{\Omega(r)}$ profiles used for stability analysis. Kep: $\Omega(r)/\Omega_0= r^{-3/2}$; Tanh: $\Omega(r)/\Omega_0=a_1$Tanh$(-r/r_1+1) +1,a_1=0.84$; Tanh2: $\Omega(r)/\Omega_0=$ $a_1$Tanh$(a_2(-r/r_1+1))+1,a_1=0.9, a_2=0.6$; mKep: $\Omega(r)/\Omega_0 = 1 / (1 + ( (r-r_2) / r_1)^{3/2})$, $\Omega(r)/\Omega_0=1$ for $r<r_2$; $r_1=1,r_2=1.5$.  }
  \label{fig:profiles}
 \vspace{-4mm}
\end{figure}

\section{Numerical solutions}
The global variation of $\Omega(r)$ (with its curvature) provides the free energy for flow-driven instabilities, as can be seen from Eqs.~\ref{eq:schrodinger}-\ref{eq:terms}. To elucidate the fundamental physics due to flow and vorticity gradients, here we examine various rotation profiles (including Keplerian, Tanh, Tanh2 and modified Keplerian forms) as shown in Fig.~\ref{fig:profiles}. These profiles are chosen to be hydrodynamically stable i.e $1<q<2$. Various rotation profiles also allow us to better understand the nature of different types of non-axisymmetric modes (local vs. global). 
Below, we first solve Eq.~\ref{eq:ode} using a complex eigenvalue shooting solver~\cite{ebrahimi2022} (benchmarked with the initial value code NIMROD~\cite{SOVINEC2004355}).
We then present global stability and effective potential analysis of $m=0$ and non-axisymmetric modes with a uniform vertical magnetic field. In this paper, we only present results for the low wave numbers ($m =1$, $k_1=\pi/4$), to capture the global modes.
\begin{figure}
    \centering
    (a)   \\
    \includegraphics[width=0.9\linewidth]{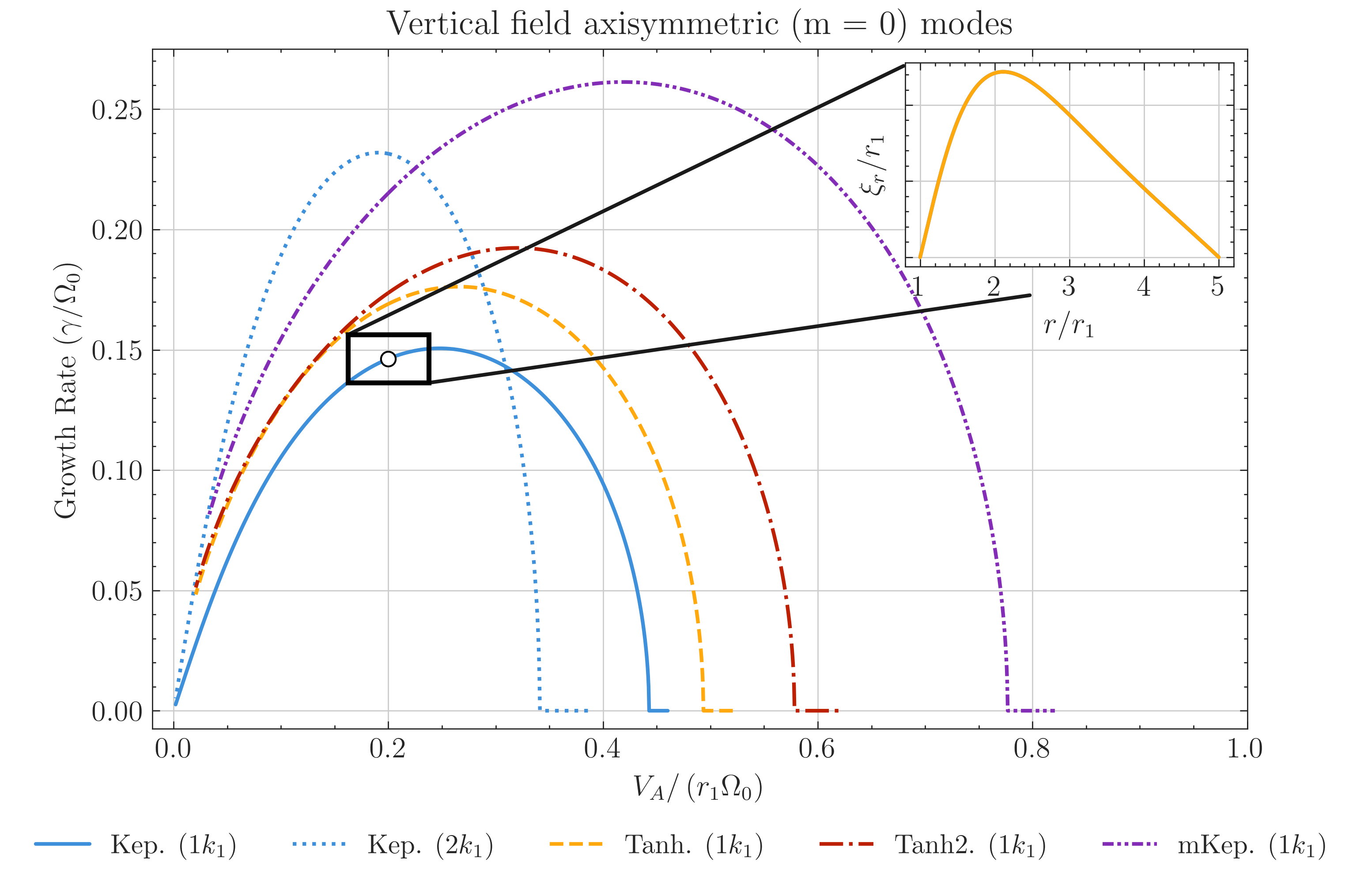}\\
    (b)\\
     \includegraphics[width=0.85\linewidth]{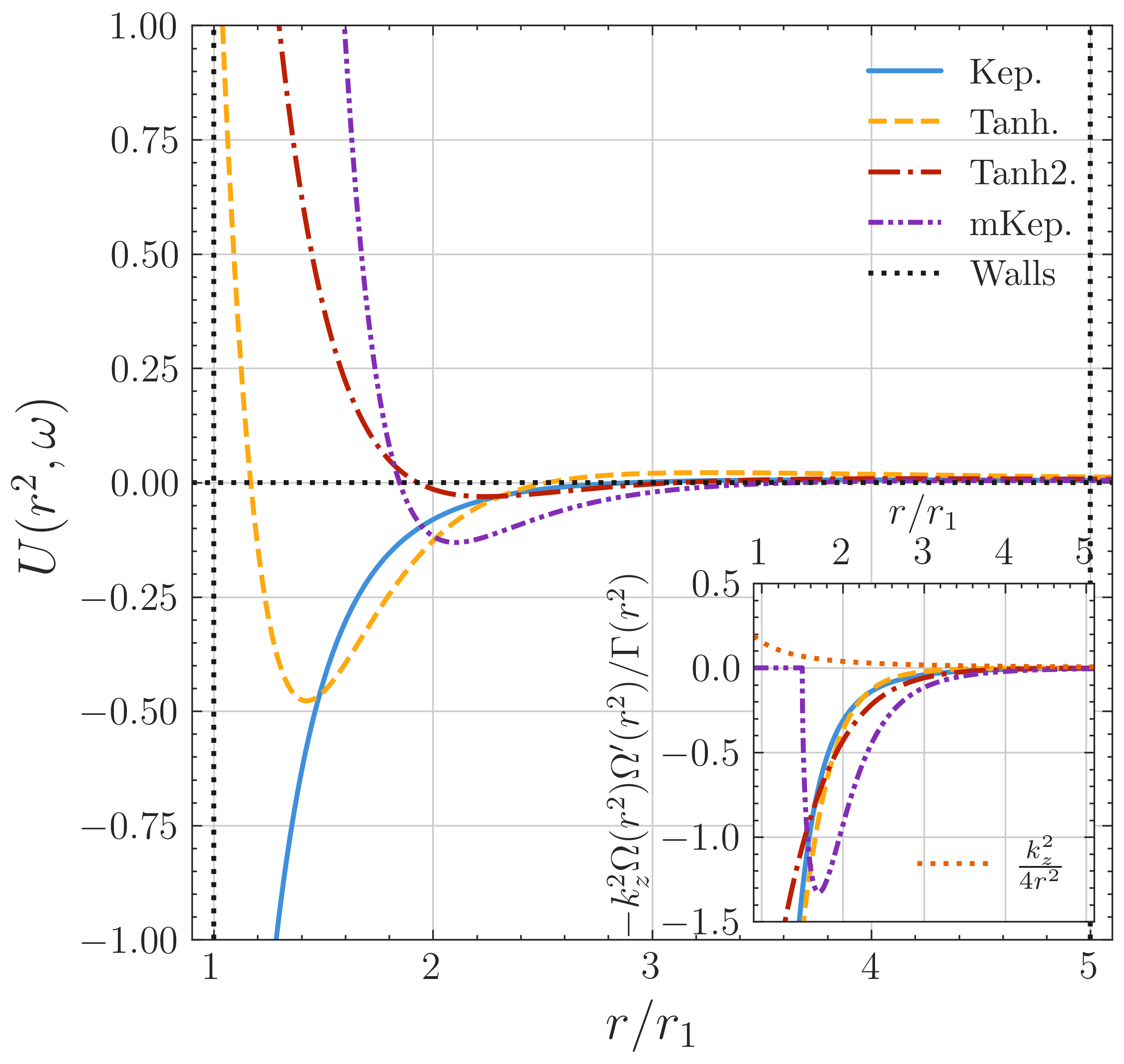}
  \caption{(a) Growth rates for $m=0$ modes (with $k=1k_1$) with global mode structure to the right, (b) effective potential profiles (for $V_A/(r_1\Omega_0)=0.2 $) for various rotation profiles in Fig.~\ref{fig:profiles}. The free energy (second term in Eq.~\ref{eq:m=0_potentia} for  marginal stability $\omega^2=0$) is shown in the lower right.  }
  \label{fig:m0growth}
 \vspace{-7mm}
\end{figure}

\subsection{Axisymmetric $m=0$ modes}

 The growth-rate values vs. magnetic-field strength for $m=0$ and $k=1k_1$ (the most global mode) for various rotation profiles is shown in Fig.~\ref{fig:m0growth}. As expected, $m=0$ MRI modes are stabilized for stronger B ($V_A/(r_1\Omega_0)$), however the instability region is extended to larger B for rotation profiles (e.g. modified Keplerian) with large global q(r) variation (Fig.~\ref{fig:profiles}(b). As mentioned above, the necessary condition for instability is when $U(r, \omega)<0$ in some region of the domain. We therefore calculate the effective potential (Eq.~\ref{eq:gpotential}) for all the rotation profiles, as shown in Fig.~\ref{fig:m0growth}(b). It is found that for flows that have more localized q(r) (or larger curvature), U(r) constitutes a negative minimum, which could result in that the mode is confined in a potential well and more localized. We note that for more localized q(r) profile of Tanh, the potential well is deep (with smaller growth rate) while for less localized q(r) profiles (Tanh2 and mKep) the well is shallow (with larger growth rate). For uniform q(r) (Keplerian flow), however, U(r) is negative all the way toward the boundary (here to the left boundary).
 \begin{figure}
\vspace{-4mm}
    \centering
    \includegraphics[width=0.8\linewidth]{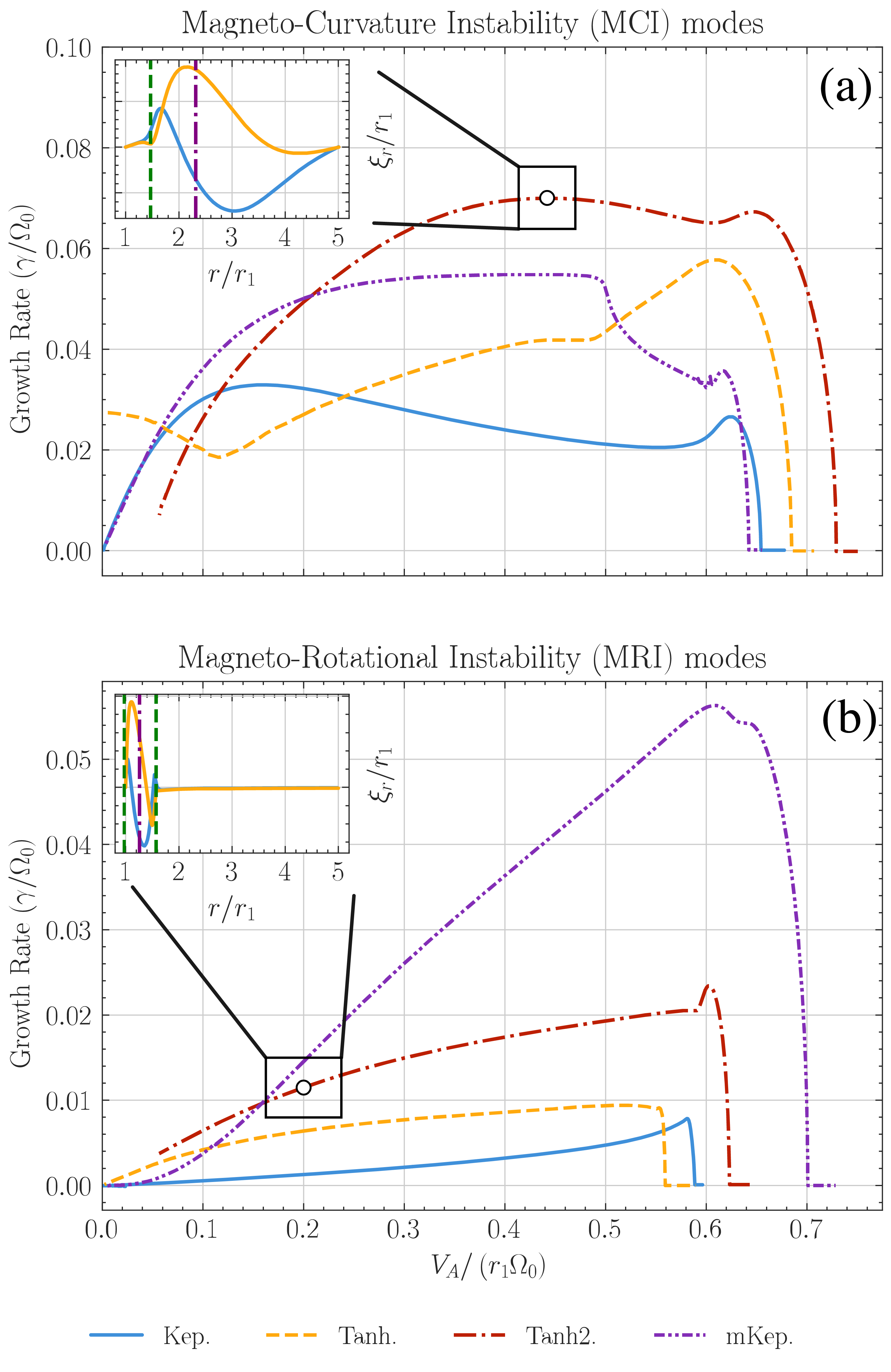}
     \includegraphics[width=2.7in]{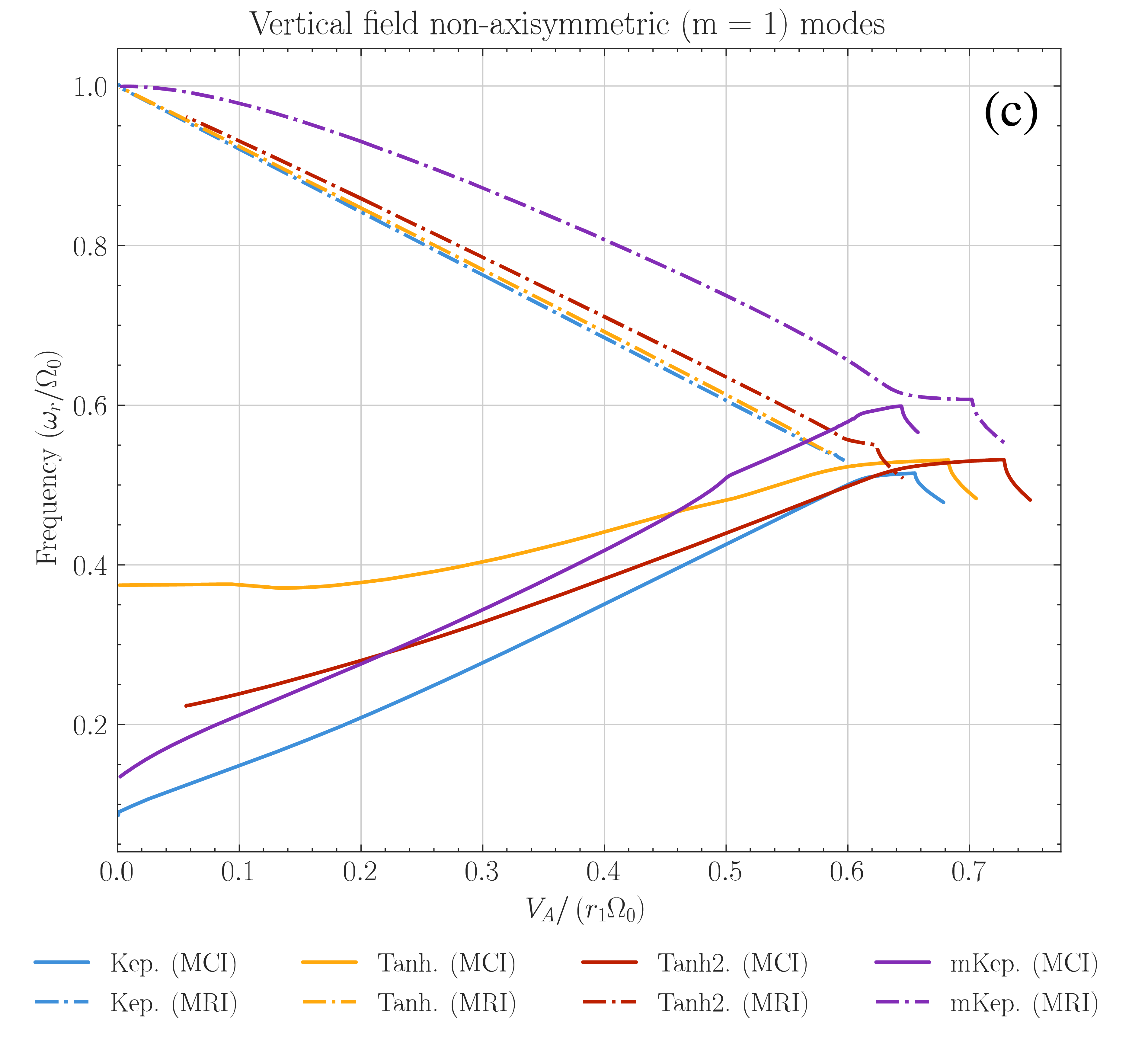}\\
  \caption{Growth rates for $m=1$ modes (with $k=1k_1$) (a) low-frequency branch MCI (global mode structure to the left) (b) high-frequency MRI (local mode structure to the left). The red and green vertical lines on the mode structures are the co-rotation (radius) and Alfvénic resonances, respectively. 
  (c) Frequencies for the two branches. }
  \label{fig:m1growth}
 \vspace{-6mm}
\end{figure}
\begin{figure}
    \centering
    \includegraphics[width=3.5in]{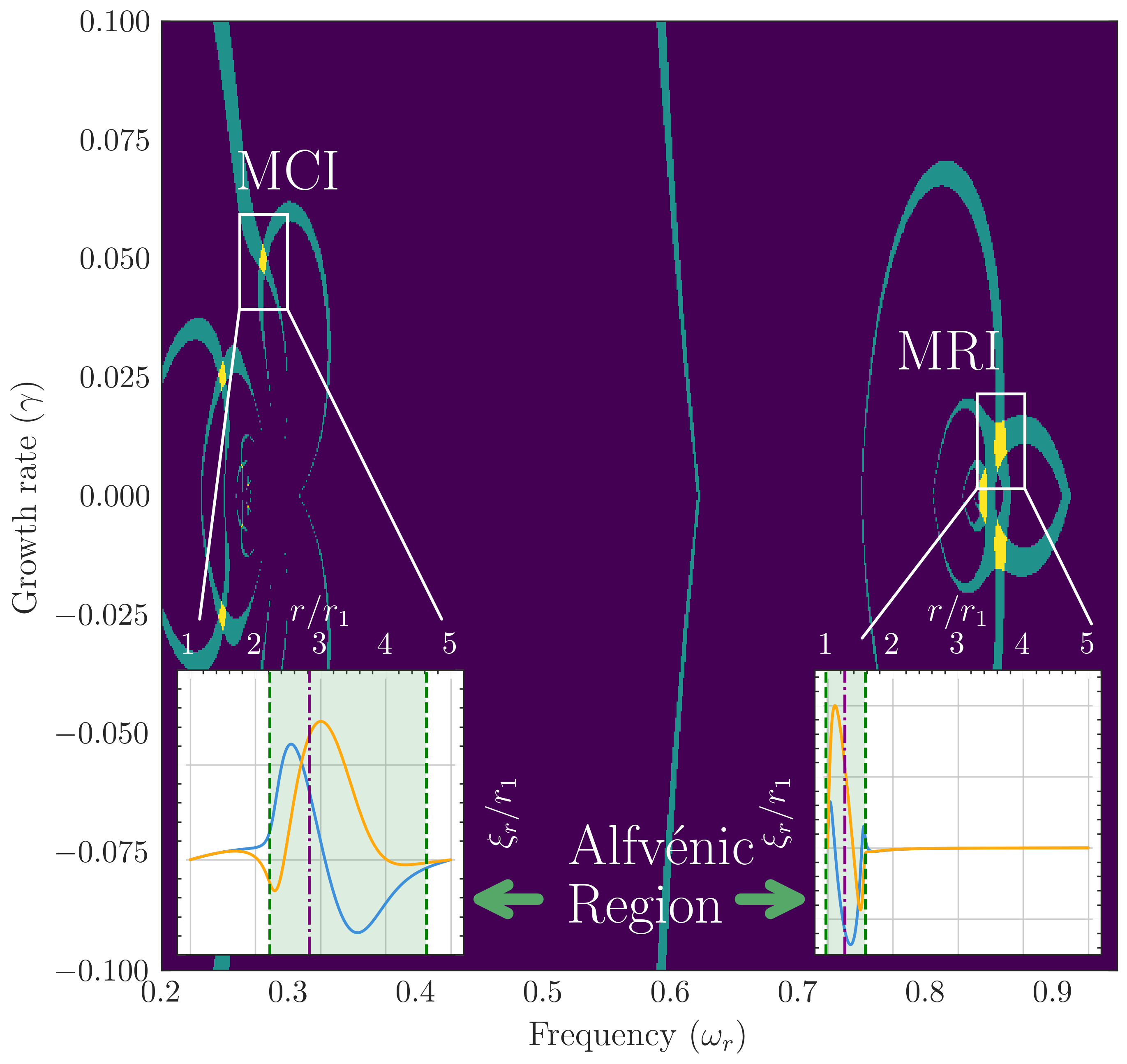}

    \caption{Scans (in a 2D plane) of the real and imaginary shooting boundary condition tails and their overlap for $m=1$, $k=1k_1$, $V_A/(r_1\Omega_0)=0.2$, for Tanh2 rotation profile. The solution (shown in yellow) exists when both tails go to zero and there is a mode. Teal indicates only one tail is near zero. The global MCI and local MRI mode structures are shown, both confined between two Alfv\'enic points (green shaded area between the green vertical lines). The vertical red line is the co-rotation point.}
    \label{fig:scan}
\end{figure}

\subsection{Two distinct non-axisymmetric modes}
\textit{Two distinct non-axisymmetric modes:} 
For non-axisymmetric $m=1$, the imaginary (growth rates) and the real (frequency) parts of the eigenvalues for various magnetic field strength, in terms of normalized Alfven velocity (the Lehnert number $B_0=V_A/V_0 = B_z/(r_1\Omega_0 \sqrt{\mu_0\rho})$), and various flow profiles are shown in Figs.~\ref{fig:m1growth}. 
 We find two sets of distinct non-axisymmetric m = 1 instabilities with different mode structures.~\cite{ebrahimi2022} The first, are the usual localized MRI modes (inner modes close to inner boundary centered about the point of maximum flow shear with high frequency); the second, which can feature more global structure at lower magnetic fields, we refer to as magneto-curvature MCI modes (outer modes close to the outer boundary). The modes are distinct in terms of frequency. 
 For weak magnetic fields, the frequency of each mode 
  reflects the angular velocity of the system at the mode's local position, as the magnetic field increases, the mode's frequency converges on a moderate frequency (Fig.~\ref{fig:m1growth}(c)). 
 
 The two distinct natures of $m=1$ modes can better be recognized in Fig.~\ref{fig:scan} when we show the complex shooting-method solution at the boundary (which we refer to as the complex and imaginary "tails") as a function of complex frequency $\omega$ for the same parameters ($m=1$, $k=1k_1$, $V_A/(r_1\Omega_0) =0.2)$. Both modes can either be confined between the two Alfv\'en resonance points ~\cite{matsumoto1995magnetic,Ogilvie_1996,ebrahimi2022,goedbloed2022} or one Alfvenic resonance and a boundary. The Alfv\'enic points 
 are the points where the magnitude of the Doppler-shifted frequency of the mode is equal to its Alfv\'en frequency, 
  [$\Re(\bar\omega)^2-\omega_A^2 =0$; $\omega_r-m\Omega=\omega_A$; $\omega_r-m\Omega = - \omega_A$], and are always centered about the point of co-rotation, where $\omega_r = m \Omega(r)$ (as shown in mode structures in Fig.~\ref{fig:scan}). 

To further elucidate the nature of non-axisymmetric modes obtained (Figs.~\ref{fig:m1growth},\ref{fig:scan}) and their associated free energies, we calculate the effective potential (Eq.~\ref{eq:gpotential}) for various rotation profiles. We find that in the Alfv\'enic propagating region where the mode is confined between the Alfv\'enic resonance points, the potential ($\Re(U)$) is negative and changes sign as it is evanescent outside this region. This can be seen for the MCI modes with Keplerian and Tanh flows (Fig.~\ref{fig:modes_potential} a-b) as well as MRI mode with Tanh2 flow shown in Fig.~\ref{fig:modes_potential}(c). For weak magnetic field, the MCI mode is confined in the outer region (Fig.~\ref{fig:modes_potential} a), and a potential well is formed between the two Alfv\'enic points and changing sign as the mode becomes evanescent on the other side of the resonances. Similarly a negative potential in the region of the confinement of the inner MRI  mode  is shown (Fig.~\ref{fig:modes_potential} d). Interestingly, for the Tanh, and Tanh2 profiles (Fig.~\ref{fig:modes_potential} d, and Fig.~\ref{fig:scan} to the left), the modes are also confined (in the Alfv\'enic region) in a potential well but right in the middle of the domain (away from the boundaries). In the hydrodynamical regimes, a similar natural confinement of non-axisymmetirc modes in a Rossby region have been demonstrated.~\cite{ono2016}Here waves are amplified between the Alfv\'enic points to cause the instability, but are strongly damped outside this region. The other sets of modes do extend between one Alfv\'enic point and a boundary and therefore can be more global. The potential ($\Re(U)$) for a global MCI mode with modified Keplerian flow, which has a shallower well and gradually passes zero as it gets closer to the boundary, is shown in Fig.~\ref{fig:modes_potential}(c). 

\begin{figure}
(a)   \hspace{37mm} (b) \\
    \includegraphics[width=0.49\columnwidth]{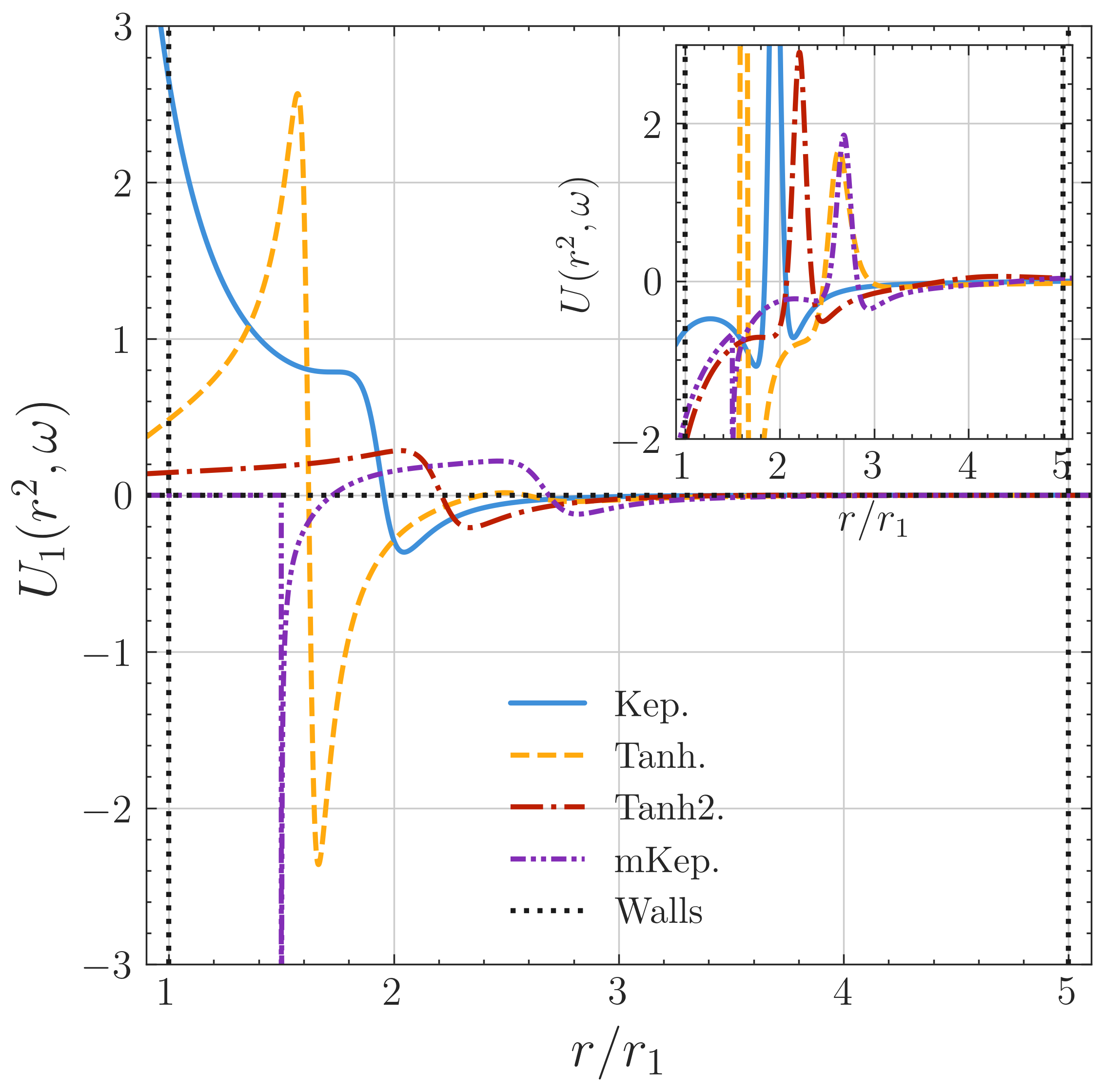} 
   \includegraphics[width=0.49\columnwidth]{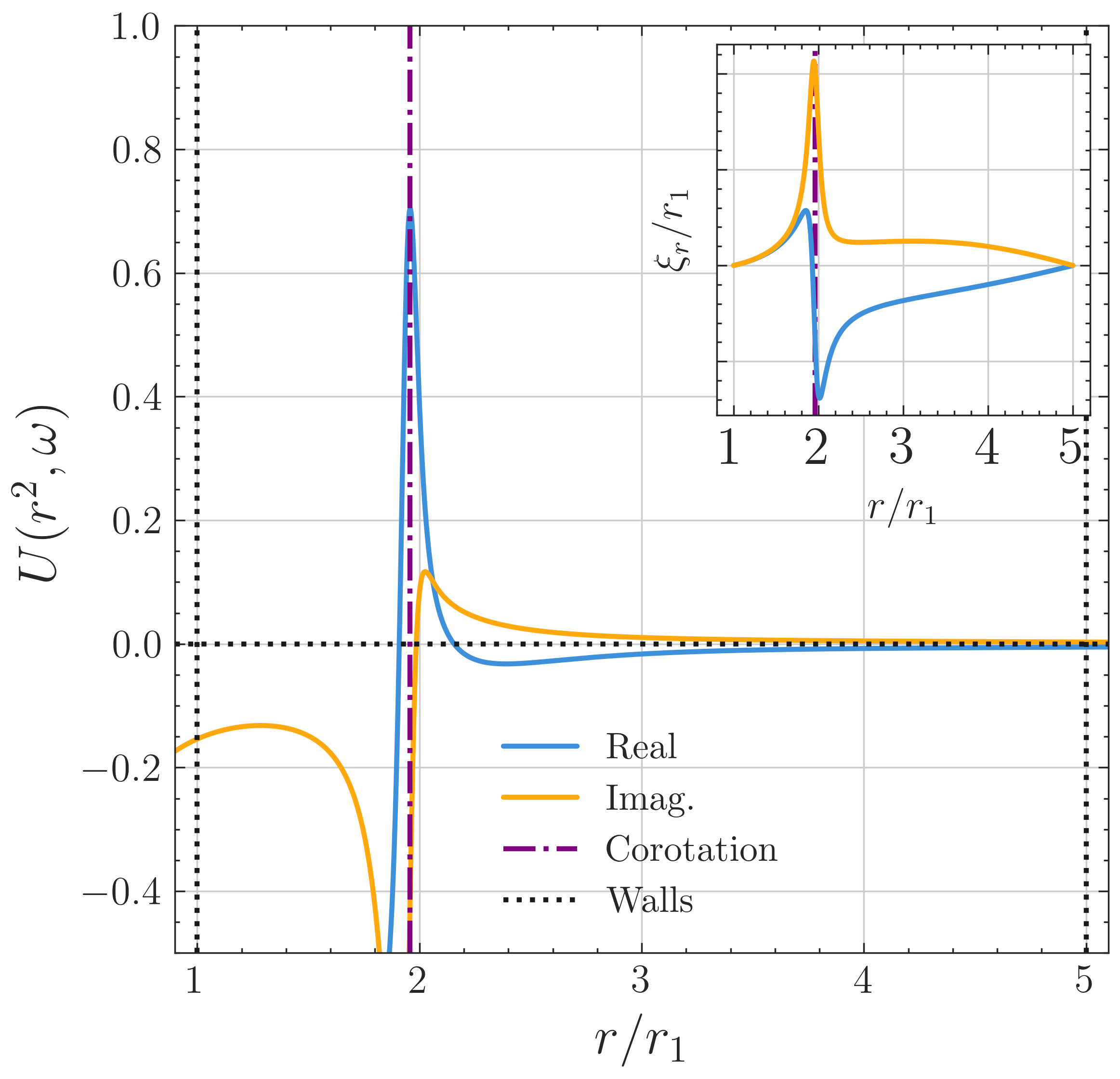}\\
    \caption{(a) The first term of potential $U_1(x,\omega, m) = \frac{\Gamma^{\prime\prime}(x)}{2\Gamma(x)}$ for all the flows ($m=1$, $k=1k_1$, $V_A/(r_1\Omega_0)=0.2$), total U is shown to the right (b) total U(x) for $m=1$, Tanh profile when B=0. }
    \label{fig:comps}
\end{figure}

Lastly, we find that all the terms in Eq.~\ref{eq:gpotential} contribute to the onset of non-axisymmetric modes, in particular the second derivative of $\Omega(r)$ can be a dominant contributor to the negative effective potential. Fig.~\ref{fig:comps}(a) shows the first term of the potential, $U_1(x,\omega, m) = \frac{\Gamma^{\prime\prime}(x)}{2\Gamma(x)}$
= $((m\Omega^{\prime})^2 + m\wb\Omega^{\prime\prime})/\Gamma(x)$ for all flows for $V_A/V_0=0.2$. It thus shows that the (curvature) second derivative of rotation profile, as well as the first derivative, does provide the free energy for the non-axisymmetric modes.  Additionally, although q(r) is less than 2 for our rotation profile (Fig.~\ref{fig:profiles}), we still find a  hydrodynamically unstable mode (at zero magnetic field), when q(r) is radially localized and has an extremum (the Tanh profile in Fig.~\ref{fig:profiles}). This results in a very localized mode at the co-rotation point and a negative potential around that point (and becomes positive close to the boundary). 

\section{Summary}
\begin{figure}
    (a) Keplerian: MCI \hspace{20mm} (b) Tanh: MCI\\
    \includegraphics[width=0.49\columnwidth]{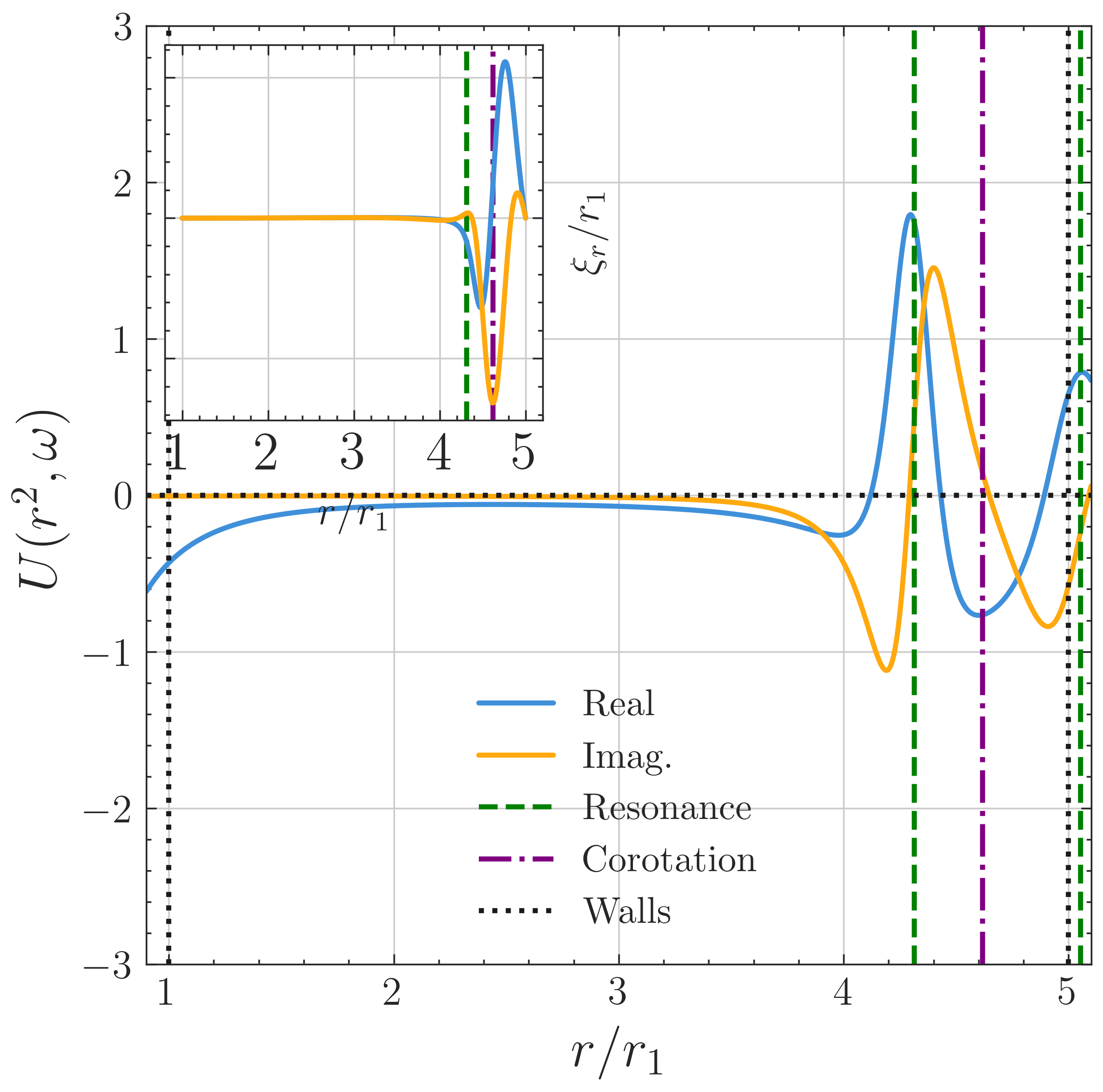} 
   \includegraphics[width=0.49\columnwidth]{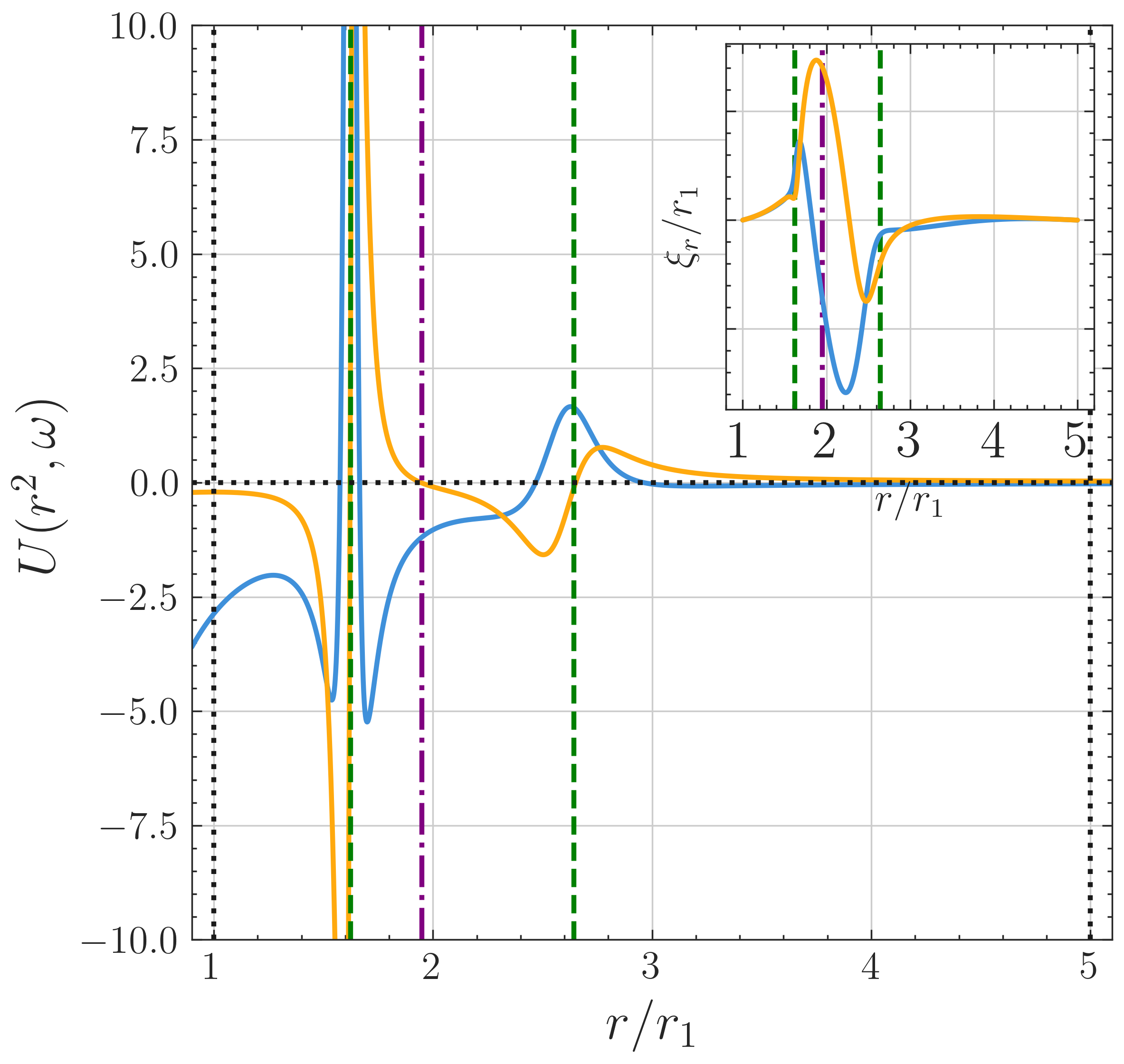}\\
    (c) MKeplerian: MCI \hspace{20mm} (d) Tanh2: MRI\\
     \includegraphics[width=0.49\columnwidth]{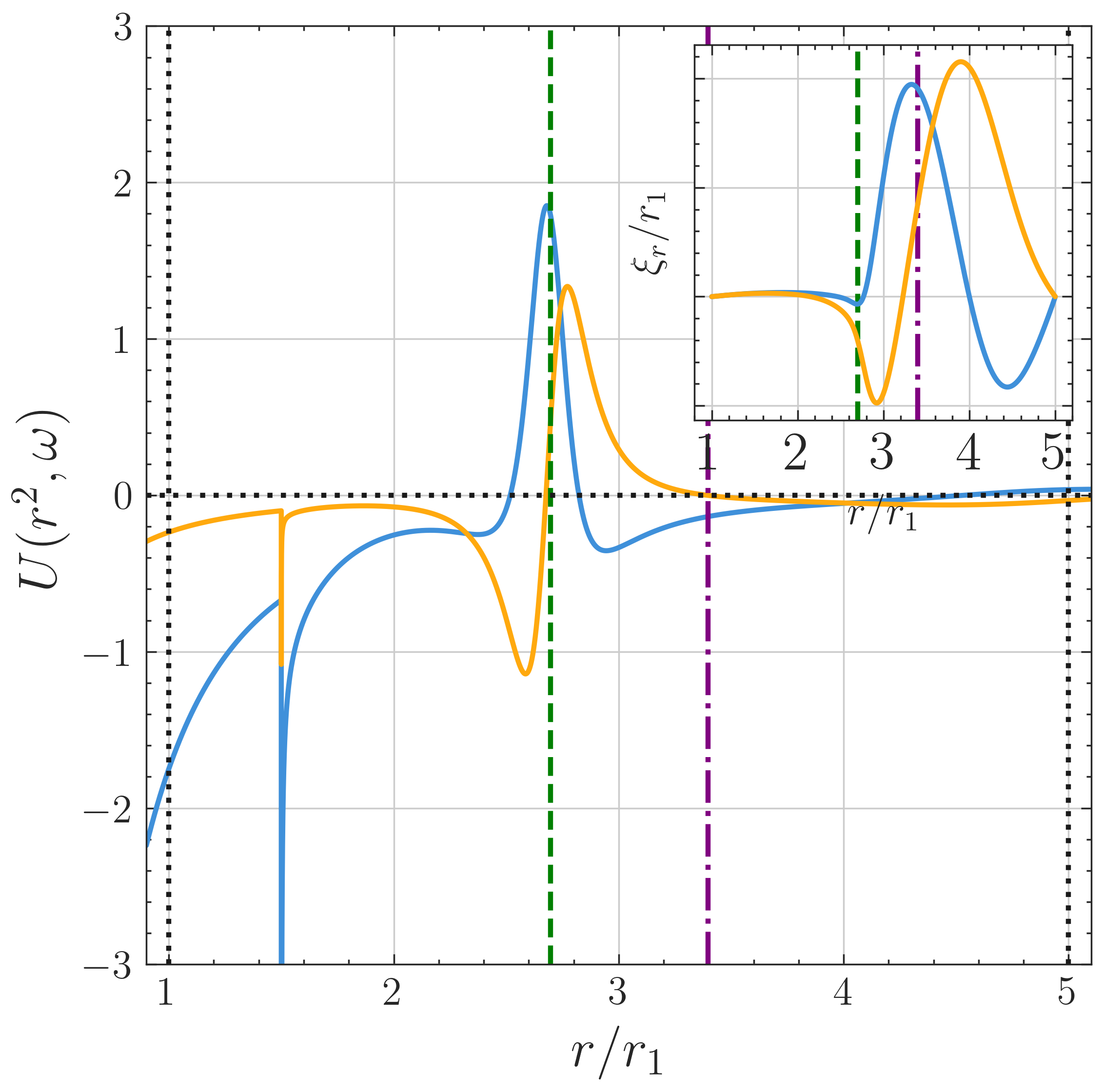}
        \includegraphics[width=0.49\columnwidth]{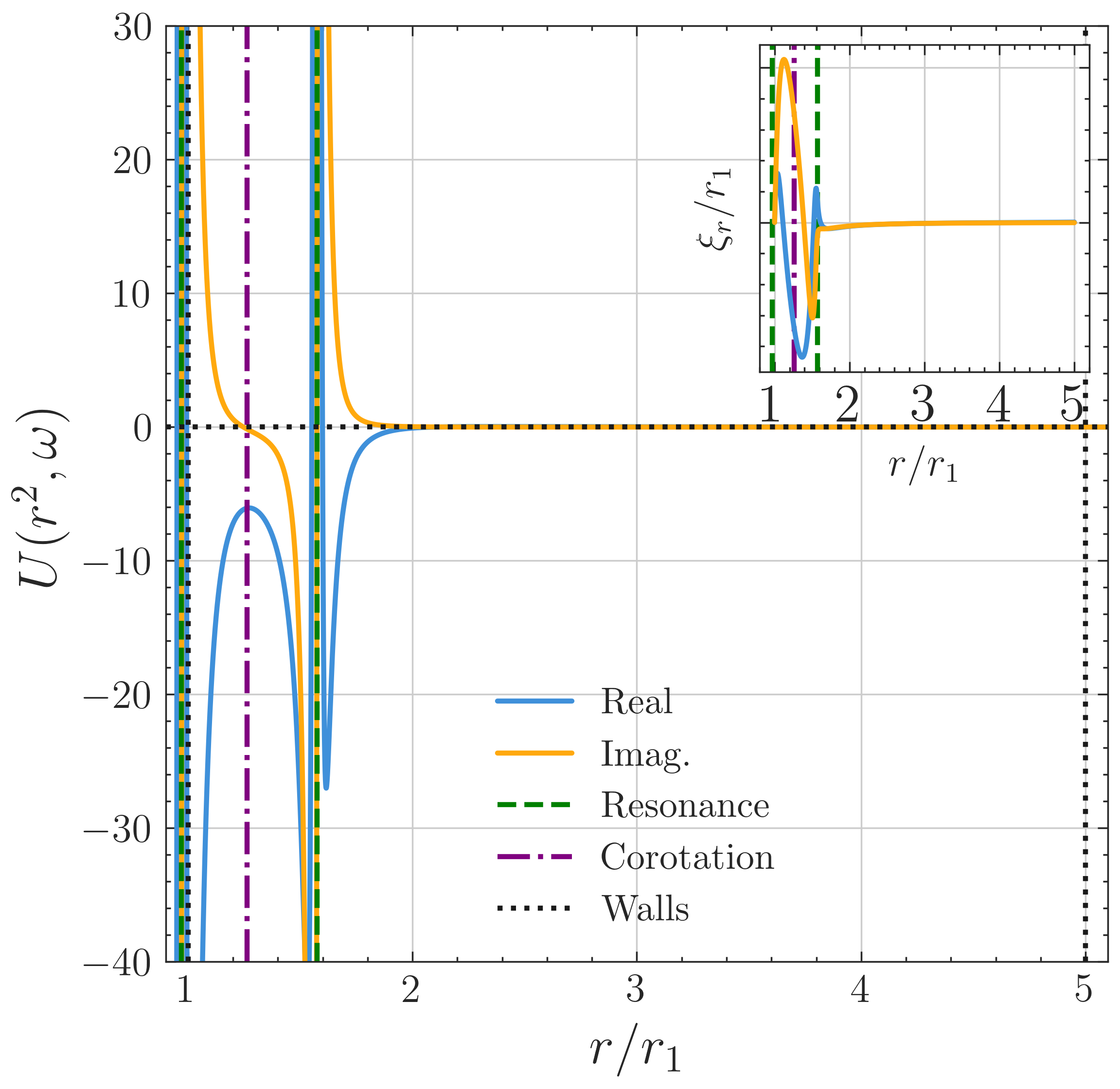}
    \caption{The effective potential (real and imaginary shown in blue and yellow) and the mode structure (in the corner) for $m=1$ and $k=1k_1$ for (a) for weak field $V_A/(r_1\Omega_0) =0.015$, Keplerian MCI (outer mode), , and for stronger field $V_A/(r_1\Omega_0) =0.2$ in (b)-(d) for various rotation profiles.}
    \label{fig:modes_potential}
    \vspace{-5mm}
\end{figure}

In summary, We have presented a generalized effective potential, which bears essential information regarding the global stability of differentially rotating systems. We have examined the stability of several rotation profiles and demonstrated that non-axisymmetric modes are triggered due to amplification of Alfv\'en waves in a potential well, which is confined between two Alfv\'enic resonant points,~\cite{matsumoto1995magnetic,Ogilvie_1996,curry1996global}  where the magnitude of the Doppler-shifted wave frequency is equal to the Alfv\'en frequency. The potential well reveals the evanescent nature of the mode outside the resonances. The form of effective potential for two branches of non-axisymmetric modes~\cite{ebrahimi2022}, a high-frequency branch (with frequency close to the inner rotation frequency of the system) and localized MRI, and a low-frequency branch (close to the rotation frequency of the outer domain), is presented.  The latter has a rather global nature due to the global rotation and its curvature, as well as global space curvature, and we call it a curvature MRI or magneto-curvature MCI mode. It is extended globally between two Alfvenic points (due the global $\Omega(r)$ and q(r)  variation, e.g. mode structure in Fig.~\ref{fig:scan} to the left) or between an Alfv\'enic point and a boundary (mode structure of modified Keplerian in Fig.~\ref{fig:modes_potential}), where the effective potential is negative but changes sign in the domain to meet the boundary.   

Here, we have focused on the ideal Alfv\'enic MHD nature of the global non-axisymmetric modes in differentially rotating systems. Non-ideal effects (including finite resistivity and viscosity) relevant to laboratory experiments~\cite{wang2022identification,wang2024observation}, and additional hydrodynamical branches will be presented in future work. 

\begin{acknowledgements}
 
This work was supported by National Science Foundation award number NSF 2308839, and DOE grant DE-AC02-09CH11466.

\end{acknowledgements}



\bibliographystyle{apsrev}



\end{document}